\documentclass[aps,prb,twocolumn,showpacs,preprintnumbers,
amsmath,amssymb,superscriptaddress]{revtex4}
\usepackage{graphicx}
\usepackage{bm}

\begin{document}

\title{Proximity effect at superconducting Sn-Bi$_2$Se$_3$ interface}

\author{Fan Yang}
\affiliation{Daniel Chee Tsui Laboratory, Beijing National
Laboratory for Condensed Matter Physics, Institute of Physics,
Chinese Academy of Sciences, Beijing 100190, China}
\author{Yue Ding}
\affiliation{Daniel Chee Tsui Laboratory, Beijing National
Laboratory for Condensed Matter Physics, Institute of Physics,
Chinese Academy of Sciences, Beijing 100190, China}
\author{Fanming Qu}
\affiliation{Daniel Chee Tsui Laboratory, Beijing National
Laboratory for Condensed Matter Physics, Institute of Physics,
Chinese Academy of Sciences, Beijing 100190, China}
\author{Jie Shen}
\affiliation{Daniel Chee Tsui Laboratory, Beijing National
Laboratory for Condensed Matter Physics, Institute of Physics,
Chinese Academy of Sciences, Beijing 100190, China}
\author{Jun Chen}
\affiliation{Daniel Chee Tsui Laboratory, Beijing National
Laboratory for Condensed Matter Physics, Institute of Physics,
Chinese Academy of Sciences, Beijing 100190, China}
\author{Zhongchao Wei}
\affiliation{Daniel Chee Tsui Laboratory, Beijing National
Laboratory for Condensed Matter Physics, Institute of Physics,
Chinese Academy of Sciences, Beijing 100190, China}
\author{Zhongqing Ji}
\affiliation{Daniel Chee Tsui Laboratory, Beijing National
Laboratory for Condensed Matter Physics, Institute of Physics,
Chinese Academy of Sciences, Beijing 100190, China}
\author{Guangtong Liu}
\affiliation{Daniel Chee Tsui Laboratory, Beijing National
Laboratory for Condensed Matter Physics, Institute of Physics,
Chinese Academy of Sciences, Beijing 100190, China}
\author{Jie Fan}
\affiliation{Daniel Chee Tsui Laboratory, Beijing National
Laboratory for Condensed Matter Physics, Institute of Physics,
Chinese Academy of Sciences, Beijing 100190, China}
\author{Changli Yang}
\affiliation{Daniel Chee Tsui Laboratory, Beijing National
Laboratory for Condensed Matter Physics, Institute of Physics,
Chinese Academy of Sciences, Beijing 100190, China}
\author{Tao Xiang}
\affiliation{Daniel Chee Tsui Laboratory, Beijing National
Laboratory for Condensed Matter Physics, Institute of Physics,
Chinese Academy of Sciences, Beijing 100190, China}
\affiliation{Institute of Theoretical Physics, Chinese Academy of
Sciences, Beijing 100190, China}
\author{Li Lu}\email[Corresponding authors: ]{lilu@iphy.ac.cn}
\affiliation{Daniel Chee Tsui Laboratory, Beijing National
Laboratory for Condensed Matter Physics, Institute of Physics,
Chinese Academy of Sciences, Beijing 100190, China}

\date{\today}

\begin{abstract}
We have investigated the conductance spectra of Sn-Bi$_2$Se$_3$
interface junctions down to 250 mK and in different magnetic fields.
A number of conductance anomalies were observed below the
superconducting transition temperature of Sn, including a small gap
different from that of Sn, and a zero-bias conductance peak growing
up at lower temperatures. We discussed the possible origins of the
smaller gap and the zero-bias conductance peak. These phenomena
support that a proximity-effect-induced chiral superconducting phase
is formed at the interface between the superconducting Sn and the
strong spin-orbit coupling material Bi$_2$Se$_3$.
\end{abstract}



\maketitle

\section{INTRODUCTION}
Due to strong spin-orbit coupling (SOC), electrons in the surface
states (SS) of a topological insulator (TI) become completely
helical, forming a new category of half metals \cite{1,2,3}. Among
many exciting features of TIs, the exotic physics at the interface
between a three-dimensional (3D) TI and an $s$-wave superconductor
is of particular interest. According to theoretical predictions,
novel superconductivity with effectively spinless $p_x+ip_y$ pairing
symmetry will be induced via proximity effect, and Majorana bound
states will emerge at the edges \cite{Bolech2007,26,29,28,27}.
Several experimental schemes have been proposed to test the
predictions, but the progresses so far reported are limited to the
observations of a supercurrent in Al-Bi$_2$Se$_3$-Al junctions
\cite{Swiss}. It is not clear whether this is because the predicted
exotic properties are suppressed by the existence of bulk states
(BS) which are present in most transport measurements. Nevertheless,
there are signs that the majority electrons are still significantly
helical in the presence of BS. For example, the magneto-resistance
of Bi$_2$Se$_3$ exhibits an unusually robust weak anti-localization
behavior \cite{22q,23,25q,24q,YYwang}, indicating the existence of a
Berry phase $\pi$ in the band structure of those electrons involved
in transport measurements. Therefore, it is possible that some of
the novel properties originally predicated for ideal TIs are still
experimentally observable even in the presence of BS.

In this Article, we report our experimental investigation on the
conductance spectra of superconductor-normal metal (S-N) interface
junctions made of Sn film and Bi$_2$Se$_3$ single crystalline flake
with BS, where Sn is a simple $s$-wave superconductor, and
Bi$_2$Se$_3$ is a typical 3D TI candidate \cite{Fangzhong}. Several
anomalies were found, including a double-gap structure that develops
below the superconducting transition temperature of Sn and a
zero-bias conductance peak grown up at lower temperatures. We will
discuss the possible origins of these phenomena, and to show that
they can be interpreted by the formation of a
proximity-effect-induced chiral superconducting phase at the
interface.

\section{\label{sec:experiment}EXPERIMENT: S\lowercase{n}-B\lowercase{i}$_2$S\lowercase{e}$_3$ JUNCTIONS}

The Bi$_2$Se$_3$ flakes used in this experiment were mechanically
exfoliated from a high quality single crystal, and those with
thickness of $\sim$100 nm were transferred to degenerate-doped Si
substrates with a 300 nm-thick SiO$_{\rm 2}$ for device fabrication.
Two Pd electrodes were firstly deposited to a selected flake. Then,
an insulating layer of heavily-overexposed PMMA photoresist with a
$1\times 1 \mu{\rm m}^2$ hole at the center was fabricated on top of
the flake. Finally, 200-nm-thick Sn electrodes were patterned and
deposited via sputtering. The device structure and measurement
configuration are illustrated in Figs. 1(b) and 1(c).
Pseudo-four-terminal measurement was performed in a $^3$He cryostat
by using lock-in amplifiers, with an ac excitation current of 1
$\mu$A at 30.9 Hz.

Determined from Hall effect measurements, the thin flakes of
Bi$_2$Se$_3$ used in this experiment have a typical carrier density
of 10$^{18}$ cm$^{-3}$ and a typical mobility of 5000 cm$^{2}/$Vs at
$T=$ 1.6 K. The Sn films deposited show a sharp superconducting
transition at $T_{\rm c}\approx$ 3.8 K and with a critical field
$H_{\rm c}$ less than 60 mT at 300 mK, indicating their high quality
in term of superconductivity, in spite of the granular morphology
[Fig. 1(c)] due to self-annealing at room temperature.

For the study of proximity effect, a clean interface and a relatively
small junction resistance are necessary. If the interfacial barrier strength is too
high, no proximity effect will occur. In order to improve
the contact between Sn and Bi$_2$Se$_3$, some of the devices were
treated with Ar ion etching in a reacting ion etching system, to
remove the possible remnant photoresist in the junction area prior
to Sn deposition (with a pressure of 100 mTorr, a power of 50 W and
for $\sim$10 s). Since Ar does not react with Bi$_2$Se$_3$, the
etching is generally a physical process. Ar etching is found to be
helpful to enhance the transparency of the junction. But good
contact can still be achieved without etching. The primary features
of the conductance spectra were found to be similar for devices with
comparable interfacial resistance, regardless of the treatment prior
to Sn deposition.

\begin{figure}
\includegraphics[width=0.84 \linewidth]{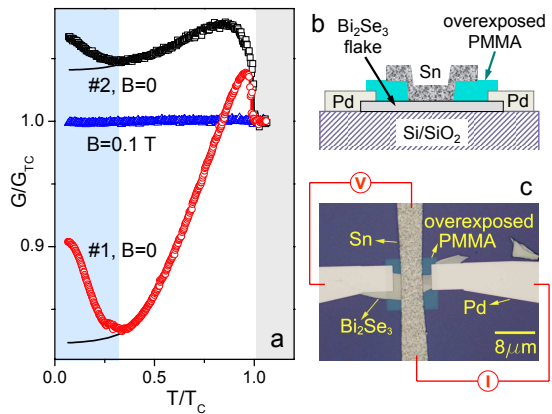}
\caption{\label{fig:fig1} {(color online) (a) Temperature
dependencies of zero-bias conductance of Sn-Bi$_{\rm 2}$Se$_{\rm 3}$
junction devices \#1 and \#2. The temperature is normalized to the
superconducting transition point of the Sn film, $T_{\rm c}\approx
3.8$ K, and the conductance is normalized to its value just above
$T_{\rm c}$. The solid lines represent the tendency of saturation
expected from BTK theory for $s$-wave S-N junctions with a
medium interfacial scattering strength \cite{30}. A magnetic field
of 0.1 T completely suppressed all the features below $T_{\rm c}$.
The white and blue regions correspond to the two stages of proximity
effect discussed in the text. (b) Illustrations of the device
structure. (c) Optical image of the device and measurement
configuration. }}
\end{figure}

More than a dozen devices were fabricated and measured at $T$=1.6 K,
six of them were further investigated down to 250 mK. All devices
exhibited qualitatively similar features. In this Article, we show
the data taken from three typical devices, labeled as \#1, \#2 and
\#3. The normal-state resistance (taken at $T=4$ K) of these devices
are 13.5 $\Omega$, 7.5 $\Omega$ and 10.2 $\Omega$, respectively.

Figure 1(a) shows the measured zero-bias differential conductance
$G$ as a function of $T$ for device \#1 and \#2, where the data are
normalized to their values above $T_{\rm c}\approx 3.8$ K. With
decreasing $T$, the conductance increases abruptly below $T_{\rm
c}$, and reaches a peak with a maximum enhancement of 3.9\% for
device \#1 and 7.7\% for device \#2, then the conductance drops
gradually until a turning point $\sim$1.2 K. Below this temperature,
the conductance increases and deviating from the saturation tendency
expected from the BTK theory \cite{30}. The deviation at 250 mK is
$\sim$8\% for devices \#1 and $\sim$3\% for devices \#2. By applying
a magnetic field $B$=0.1 T, all the low-temperature structures on
the $G-T$ curves were removed, indicating that they are closely
related to the superconductivity of Sn.

In BTK theory \cite{30}, the normalized zero-bias $dI/dV$ of an S-N
junction, $Y$, can be written as:
\begin{widetext}
$$Y(Z,T)=\left.\frac{I_{NS}(Z,T)}{I_{NN}(Z)}\right|_{eV\rightarrow0}=(1+Z^{2})\int_{-\infty}^{\infty}\left(\frac{\partial f_{0}}{\partial E}\right)[2A(E)+C(E)+D(E)]$$
\end{widetext}
where the dimensionless parameter $Z$ describes the barrier strength
of the S-N junction, $f_{0}(E,T)$ is the Fermi distribution, $A(E)$,
$C(E)$ and $D(E)$ are functions defined in the BTK theory. At $T=0$
this equation can be simplified to:
$$\left.Y\right|_{T=0}=\frac{2(1+Z^{2})}{~(1+2Z^{2})^{2}}$$
With this formula, one can estimate the $Z$ value of a device using
its saturated conductance at low temperatures.

In Fig. 1(a), the normalized $dI/dV$ for devices \#1 and \#2
saturate to about 0.82 and 1.04, as indicated by the solid lines,
which yield barrier strengths of $Z=$0.66 and 0.54 for devices \#1
and \#2, respectively. Such barriers are not in the transparent
limit (i.e., $Z=0$) nor in the tunneling limit ($Z\gg$1), which
ensures the happening of reasonably strong proximity effect at the
interface on one hand, and enabling us to probe the information of
the density of states on the other hand.

\begin{figure}
\includegraphics[width=0.88 \linewidth]{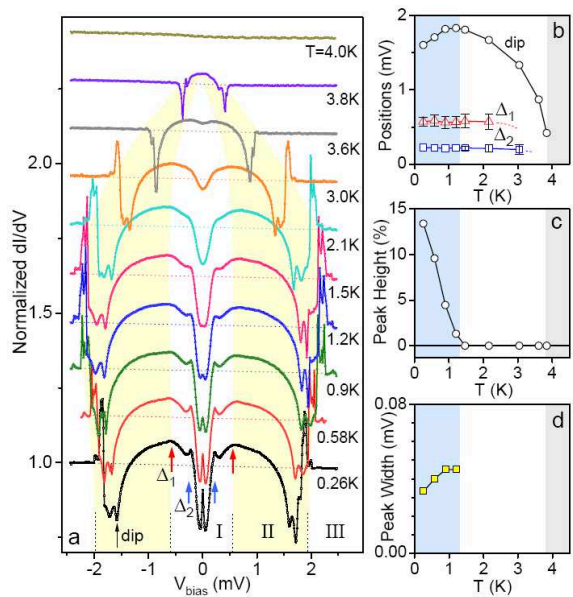}
\caption{\label{fig:fig2} {(color online) (a) Conductance spectra of
device \#1 at different temperatures. Each curve is normalized to
its high bias value, and is shifted vertically for clarity. (b)
Temperature dependencies of the positions of the dip, the first gap
$\Delta_1$, and the second gap $\Delta_2$. (c) Temperature
dependence of the peak height. The peak grows up at a temperature
significantly lower than the $T_{\rm c}$ of Sn. (d) Temperature
dependence of the full width of the peak at half height. }}
\end{figure}

In Fig. 2 and Fig. 3 we show the conductance spectra, namely the
bias voltage ($V_{\rm bias}$) dependence of differential
conductance, of device \#1 measured at different temperatures and in
different magnetic fields. Each curve is normalized to its high bias
value in region III. Three unusual features were observed, as
elaborated below.

The first feature is a bump-like enhancement, together with sharp
dips at the two sides. It develops at temperatures immediately below
$T_{\rm c}$ in regions I and II, as marked in Fig. (2), and is best
seen at high $T$ when the gaps are largely undeveloped. It
corresponds to the abrupt increase of conductance in the $G-T$
curves just below $T_{\rm c}$. With decreasing $T$ the bump
structure evolves and extends to a bias voltage several times larger
than the superconducting gap of Sn. In the mean time gap structures
develop around zero bias voltage, which will be discussed later.

\begin{figure}
\includegraphics[width=0.88 \linewidth]{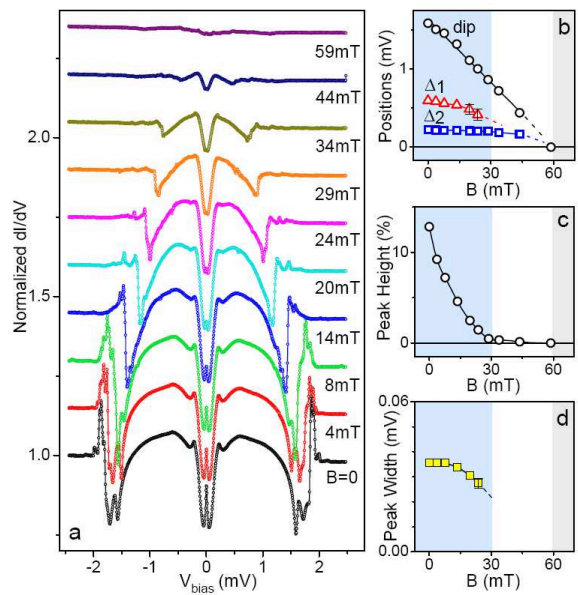}
\caption{\label{fig:fig3n3} {(color online) (a) Conductance spectra
of device \#1 taken at 300 mK and in different magnetic fields.
Curves have been shifted vertically for clarity. (b) Field
dependencies of the positions of the dip, the first gap $\Delta_1$,
and the second gap $\Delta_2$. (c) Field dependence of the peak
height. (d) Field dependence of the full width of the peak at half
height. }}
\end{figure}

The position of the dips is device-dependent. It is very close to
zero bias voltage at temperatures just blow $T_c$, but can reach as
high as 8 mV at low temperatures for high resistance junctions in
this experiment (data not shown). The dips can be safely attributed
to current-driven destruction of superconductivity in the local Sn
film surrounding the junction. We note that the use of large
measurement current (up to 200 $\mu$A here) is unavoidable in
proximity effect studies where the junctions usually need to be
reasonably transparent. Estimation shows that the local current
density at the step edge of the window of our junctions could reach
to $\sim 10^4$ to $10^5$ A/cm$^{2}$ at $V_{\rm bias}$=2 mV (for a 10
$\Omega$ junction), which may well exceed the critical current
density of the Sn film. The non-monotonic $T$ dependence of the dip
position, as shown in Fig. 2, Fig. 4 for devices \#1 and \#3, and in
Fig. 7 for comparative Sn-graphite devices, suggests that the
detailed destruction process might involve local heating and thermal
conduction which has a significant temperature dependence below
$\sim$1 K (see Section \ref{sec:comparative}).

The second feature in the conductance spectra is a double-gap
structure that develops on the enhanced conductance background, as
shown in region I of Fig. 2(a). The borders of the two gaps are
indicated by the arrows. The development of this structure is
responsible for the drop of $G-T$ curves below the peak temperatures
in Fig. 1(a). For device \#1, the first (bigger) gap is
$\Delta_1$=0.59 mV, which matches with the superconducting gap of
Sn, and the second (smaller) gap is $\Delta_2$=0.21 mV, only about
1/3 of the first one. It should be noted that for most devices (10
out of 12) only the smaller gap was clearly observed. In Figs. 4(a)
and 4(d) we show two such examples observed on devices \#3 and \#2,
respectively. For device \#3, a faint structure can still be
resolved at the $\Delta_1$ position, as indicated by the blue arrows
in Fig. 4(a), and is best seen on the curve taken at 1.2 K.

\begin{figure}
\includegraphics[width=0.88 \linewidth]{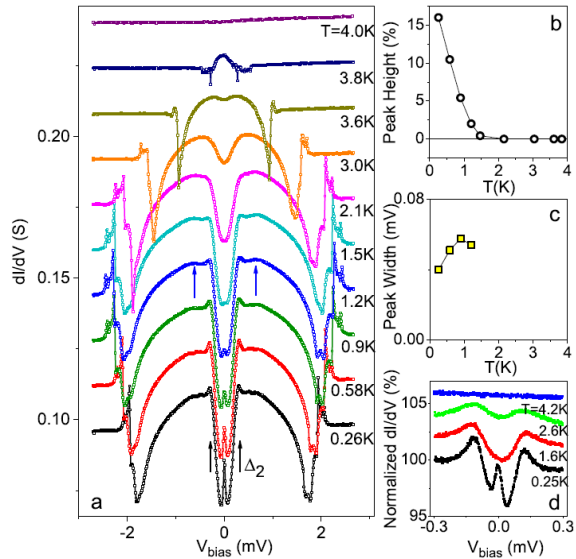}
\caption{\label{fig:fig4} {(color online) (a) Conductance spectra of
device \#3 taken in zero magnetic field and at different
temperatures. The blue arrows indicate where the coherence peak of
the bigger gap $\Delta_1$ is expected. Only a faint structure at
this position can be resolved (which is best seen on the 1.2 K
curve). Curves have been shifted vertically for clarity. (b)
Temperature dependence of the peak height at zero bias voltage. (c)
Temperature dependence of the full width of the peak at half height.
(d) Conductance spectra of device \#2 at different temperatures.
Curves are shifted vertically for clarity.}}
\end{figure}

The existence of two distinct gaps clearly indicates that the
smaller gap is not the superconducting gap of Sn. It should arise
from some new superconducting phase formed at the interface, which
will be further discussed later.

The third feature in the conductance spectra is a zero bias
conductance peak (ZBCP) developed at low temperatures. This ZBCP is
responsible for the conductance increment on $G-T$ curves below
$\sim$1.2 K. The peak height grows up almost linearly with
decreasing $T$. It reaches 13.4\% and 16.4\% of the normal-state
conductance at the lowest temperature of this experiment, 250 mK,
for device \#1 and \#3 respectively, as shown in Figs. 2(c) and
4(b). And the peak width \cite{Note_ZBCP_width} decreases with
decreasing $T$ at low temperatures, as shown in Figs. 2(d) and 4(c).
Both the height and the width of the ZBCP can be suppressed by
applying a magnetic field, as shown in Figs. 3(c) and 3(d).

The temperature dependence of peak width contains important
information about the origin of the ZBCP. When we plot the ZBCP
against bias current instead of bias voltage, as shown in Figs. 5(a)
and 5(b) for devices \#1 and \#3 respectively, the peak width
decreases with decreasing $T$ as well [Fig. 5(c)]. It indicates that
the ZBCP is originated from some kind of resonance whose width is
controlled by thermal broadening.

\begin{figure}
\includegraphics[width=1 \linewidth]{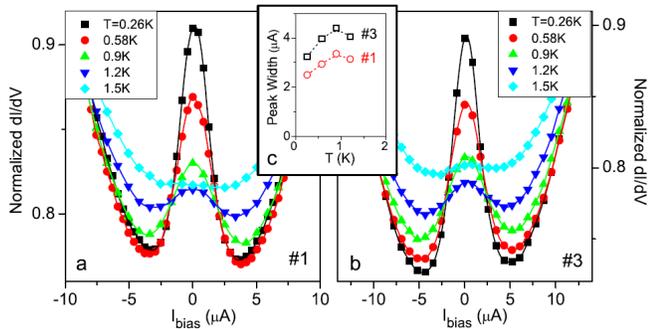}
\caption{\label{fig:fig5} {(color online) Normalized dI/dV plotted
against bias current for devices \#1 (a) and \#3 (b). (c)
Temperature dependence of the full width of the peak at half height,
for both devices \#1 and \#3.}}
\end{figure}

\section{\label{sec:comparative}Comparative Experiment: S\lowercase{n}-Graphite JUNCTIONS}
In order to examine whether the conductance anomalies observed in
Sn-Bi$_{\rm 2}$Se$_{\rm 3}$ interfacial junctions are intrinsic
properties of the interface between an $s$-wave superconductor Sn
and a helical metal, we have made two more devices for comparison by
replacing Bi$_{\rm 2}$Se$_{\rm 3}$ with bulk graphite, while keeping
other parameters the same. We choose graphite because it is similar
to Bi$_{\rm 2}$Se$_{\rm 3}$ in carrier density and mobility, but
with much weaker spin-orbit coupling strength \cite{yao}. The two
devices, made of graphite flakes of thickness $\sim$ 100 nm, are
labeled as S1 and S2, respectively. Their optical images are shown
in Fig. 6.

\begin{figure}
\includegraphics[width=0.8 \linewidth]{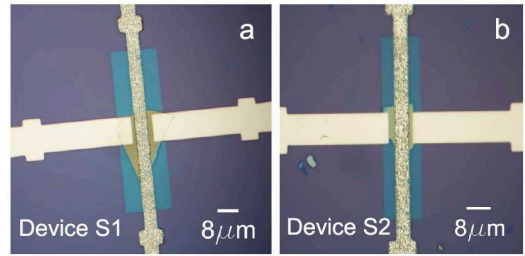}
\caption{\label{fig:fig6} { (color online) Optical images of
Sn-graphite devices S1 (a) and S2 (b).}}
\end{figure}

\begin{figure}
\includegraphics[width=0.98 \linewidth]{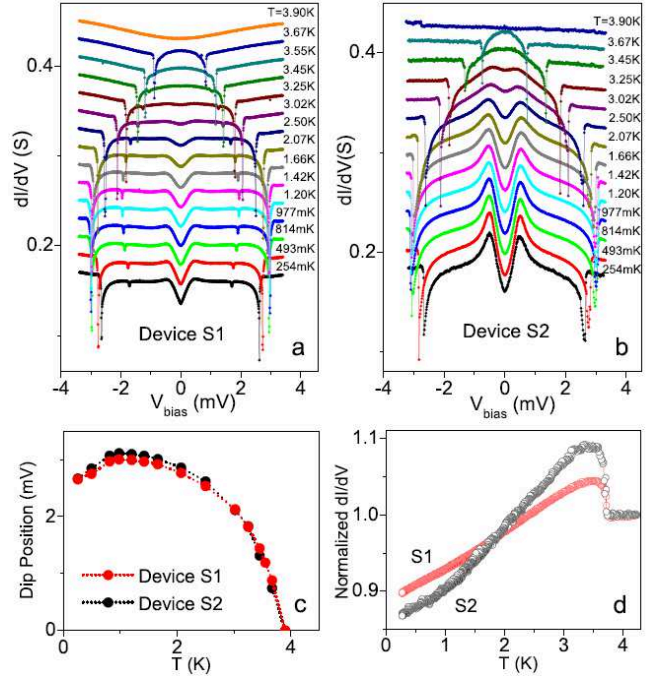}
\caption{\label{fig:fig7} { (color online) Conductance spectra of
Sn-graphite devices S1 (a) and S2 (b) taken at different
temperatures. Curves are shifted vertically for clarity. (c)
Temperature dependencies of the dip positions. (d) Normalized zero
bias $dI/dV$ as a function of temperature.}}
\end{figure}

In Fig. 7(d) we show the measured zero-bias differential conductance
$G$ as a function of temperature for these two devices. The data are
normalized to their values above $T_{\rm c} \approx 3.8$ K (around 6
$\Omega$ for both devices). The $G-T$ curves are similar to that of
the Sn-Bi$_{\rm 2}$Se$_{\rm 3}$ devices, except that there is no
upturns below $\sim$1.2 K. The conductance decreases with decreasing
temperature monotonously down to 250 mK without saturation.

Figures 7(b) and 7(c) show the conductance spectra of device S1 and
S2, respectively, taken at different temperatures. A single gap is
developed at low temperatures. The coherence peak of that gap is
located at $\sim$0.5 mV, which is close to the superconducting gap
of the Sn electrode. Neither a zero bias peak nor a second smaller
gap was observed. This single gap structure can be well understood
within the BTK theory.

There is a sharp dip at each side of the conductance spectrum,
beyond which the curve becomes flat and featureless. As discussed in
Section \ref{sec:experiment}, we attribute the dip structure to the
local destruction of superconductivity of the Sn film near the
junction. The fact that similar dips were found in both Sn-Bi$_{\rm
2}$Se$_{\rm 3}$ and Sn-graphite junctions suggests that these dips
are not specifically related to Bi$_{\rm 2}$Se$_{\rm 3}$.

\begin{figure}
\includegraphics[width=1 \linewidth]{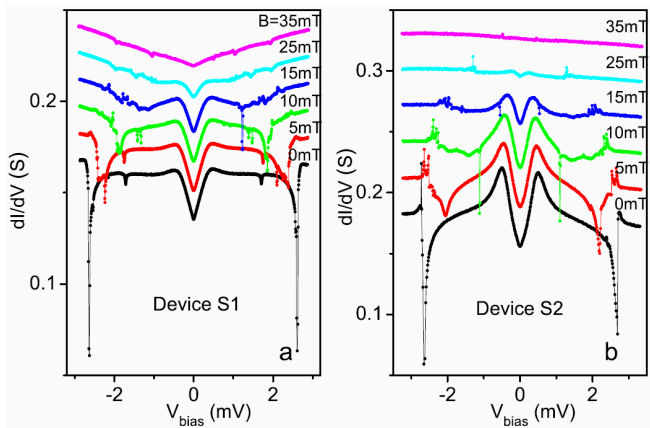}
\caption{\label{fig:fig8} { (color online) Conductance spectra of
Sn-graphite devices S1 (a) and S2 (b) taken at 250 mK and in
different magnetic fields. Curves are shifted vertically for
clarity.}}
\end{figure}

The magnetic field dependencies of the conductance spectra of
devices S1 and S2 are plotted in Figs. 8(a) and 8(b), respectively.
All the structures in the spectrum can be removed by applying a
magnetic field higher than the $H_{\rm c}$ of Sn, indicating that
they are related to the superconductivity of Sn electrode.

In summery, our comparative experiment on Sn-graphite devices
reveals only a single-gapped structure, neither a second smaller gap
nor a ZBCP was observable, unlike those on Sn-Bi$_{\rm 2}$Se$_{\rm
3}$ devices. The data also show that the non-monotonous temperature
dependence of the dip position is irrelevant to the use of Bi$_{\rm
2}$Se$_{\rm 3}$.

\section{Discussion: Possible Origins of the ZBCP and the double-gap structure}
The general trends of the $G-T$ and $G-V_{\rm bias}$ curves of the
Sn-Bi$_2$Se$_3$ devices can be understood within the framework of
the BTK theory \cite{30}, which was developed in the early 1980's to
describe the two-particle process at S-N interfaces. However, this
theory cannot explain the appearance of a second gap, nor the ZBCP
at low temperatures.

Previously, ZBCPs were also observed in some S-N \cite{a,b,c,d,f}
and S-insulator-N \cite{e} junctions, and were explained in several
different mechanisms.

The first possible mechanism is related to incoherent accumulation
of Andreev reflections (AR), which happens when there is a large
probability of backscattering due to, e.g, the involvement of the
other surface of the normal-metal thin film \cite{b,c}. ZBCPs of
this kind usually grows up immediately below $T_c$. The ZBCP
observed here seems irrelevant to this mechanism, since it has
sensitive $T$ and $B$ dependencies, appearing at much lower
temperatures.

The second possible mechanism is related to coherent scattering of
carriers near the interface due to phase conjugation between the
electron's and the hole's trajectories, leading to an enhanced AR
probability \cite{i}. This mechanism is also expressed in a more
general way by using a random matrix theory \cite{h}. A ZBCP caused
by this mechanism is sensitive to both temperature and magnetic
field, since it involves a coherent loop. However, these theories do
not take account of the strong SOC and its resulting Berry phase. In
the presence of strong SOC, the phase accumulated by the incident
electron along its path cannot be canceled by the retro-reflected
hole, i.e., the phase difference between the $N^{\rm th}$ and the $
(N+1)^{\rm th}$ reflected hole is not zero. Furthermore, the theory
in Ref. [27]
suggests that this kind of ZBCP often appear in
junctions with relatively strong scattering rate, and that the value
of the conductance peak will not exceed the conductance of the
normal state, whereas in our experiment the ZBCP can be higher than
the conductance of the normal state, as shown in Fig. 1(a) for
device \#2. Therefore, we believe that the ZBCP observed in our
experiments is not caused by the aforementioned constructive
interference.

The third explanation is to phenomenologically attribute ZBCP to a
pair current flowing between the superconducting electrode and the
proximity-induced superconducting phase \cite{a}. For a ZBCP of this
type, its behavior will resemble the critical supercurrent of a
Josephson junction. As temperature decreases, the critical current
of a Josephson junction will first increase then get saturated. For
a ZBCP of this type, therefore, its peak width is expected to
increase with decreasing $T$ if it is plotted against bias current.
However, the ZBCP in this experiment shrinks with decreasing $T$, as
can be seen in Fig. 5. Therefore, the pair current picture seems
inapplicable to our results.

Another possible mechanism of ZBCP involves unconventional
superconductivity with an asymmetric orbital order parameter
\cite{SrRuO,5,4,CR_Hu,6,7,8,9}. For example, in the $p$-wave
superconductor Sr$_2$RuO$_4$, the reflection of order parameter at
the S-N edge in the $ab$-plane feels a sign-change, giving rise to
an Andreev bound state at the Fermi energy and a ZBCP in tunneling
measurement \cite{SrRuO}. After having ruled out other possible
mechanisms to the best of our knowledge, we believe that this
mechanism involving unconventional superconductivity is most likely
responsible for the appearance of the ZBCP in this experiment.

Our entire picture for the observed phenomena is as follows. With
decreasing $T$ to below $T_{\rm c}$ of Sn, proximity effect develops
at the S-N interface via two-particle exchange processes, i.e.,
Cooper pairs are exchanged from the Sn side to the Bi$_2$Se$_3$
side, and entangled quasi-particle pairs are exchanged back in a
time-reversal process, known as AR. Unlike in an usual proximity
effect, where only a single gap of the parent superconductor is
seen, the observation of a second gap here indicates the formation
of a new S'-N interface where S' is not Sn but a
proximity-effect-induced superconducting Bi$_2$Se$_3$ phase. The
original Sn-Bi$_2$Se$_3$ interface becomes an S-S' interface whose
role vanishes in resistance measurement, so that the gap structure
related to Sn is either absent or largely suppressed. We speculate
that the suppression of back-scattering due to strong SOC and the
high electron mobility in Bi$_2$Se$_3$ would help maintaining the
coherence of the two-particle AR process in space and time domains,
thus stabilizing the proximity-effect-induced new superconducting
phase (S'-phase) in a substantially large volume of Bi$_2$Se$_3$.

With further decreasing $T$, the new superconducting phase becomes
uniformly developed, then the coherence peaks at the two shoulders
of the gap appear. The ZBCP grows up simultaneously, presumably also
related to the improved uniformity of the new superconducting phase
at low temperatures.

Since no ZBCP is observed in similar devices made of graphite which
has negligible SOC, one would naturally speculate that strong SOC is
the cause for generating the ZBCP in Sn-Bi$_2$Se$_3$ devices. The
important role of strong SOC to the electron states in Bi$_2$Se$_3$
has been revealed by the unusually robust behavior of electron
weak-anti-localization \cite{22q,23,25q,24q,YYwang}, a phenomenon
which mainly occurs in 2D electron systems. It indicates the
existence of a Berry phase $\pi$ in the electrons' band structure.
The new superconducting phase inhabited in such a background, with
inter-locked momentum and spin degrees of freedom, is believed to
own effectively a spinless $p_x+ip_y$ pairing symmetry. The
asymmetric orbital part of the order parameter, inherited from the
Berry curvature of the bands, forms resonant bound states at the
S'-N interface due to the interference between the incoming and
reflecting waveforms there, hence giving rise to the observed ZBCP.
With such a picture, it is natural that the peak width gets narrower
with reduced thermal broadening at lower $T$.

\section{Conclusion}
To summarize, we have investigated the conductance spectra of S-N
junctions between an $s$-wave superconductor Sn and a strong SOC
material Bi$_2$Se$_3$. A small gap different from that of Sn was
clearly resolved, together with a ZBCP growing up at low
temperatures. The results indicate the formation of a new
superconducting phase with unconventional pairing symmetry at the
interface. Our work would encourage future experiments to search for
Majorana fermions and other pertinent properties by employing hybrid
structures of $s$-wave superconductor and topological
insulator-related materials.

\begin{acknowledgments}
We would like to thank H. F. Yang and C. Q. Jin for experimental
assistance, L. Fu, Z. Fang, X. Dai, Q. F. Sun, X. C. Xie and S. C.
Zhang for stimulative discussions. This work was supported by the
National Basic Research Program of China from the MOST under the
contract No. 2009CB929101 and 2011CB921702, by the NSFC under the
contract No. 11174340 and 11174357, and by the Knowledge Innovation
Project and the Instrument Developing Project of CAS.

\vspace{0.5cm}

\noindent \textit{Note added in reversion}: After the submission of
this manuscript, observation of a supercurrent and possible evidence
of Pearl vortices were reported in W-Bi$_2$Se$_3$-W junction
\cite{Pennsylvania}, a ZBCP was observed on normal metal-Cu$_{\rm
x}$Bi$_2$Se$_3$ point contact \cite{Ando}, and evidence of perfect
Andreev reflection of the helical mode was obtained in InAs/GaSb
Quantum Wells \cite{RRDu}.
\end{acknowledgments}


\begin{thebibliography}{10}
\bibitem{Note_ZBCP_width}  There is about $10\ \mu$V of smearing on the ZBCP by the small ac voltage used
in the measurement.
\end{thebibliography}
\end{document}